\documentclass{article}
\usepackage{eurosym}
\usepackage{amssymb}
\usepackage{amsmath,graphicx}
\usepackage[margin=1in]{geometry}
\usepackage{xcolor}
\usepackage{hyperref}

\begin{document}

\title{Whip dynamics in a freely falling chain}
\author{Tianyi Guo$^1$, Xiaoyu Zheng$^2$, Peter Palffy-Muhoray$^{1,2}$ \\
\emph{$^1$Advanced Materials and Liquid Crystal Institute, Kent State
University, OH, USA}\\
\emph{$^2$Department of Mathematical Sciences, Kent State University, OH,
USA }}
\maketitle

\begin{abstract}
In this work, we examine the whip dynamics in a freely falling chain. We
consider an inextensible chain with free ends in the presence of gravity
hanging from a fixed pulley. If the configuration is unbalanced, the chain
begins to accelerate and eventually may lift off the pulley. We show that
after the liftoff, the dynamics of the chain resembles that of a whip whose
tip speed diverges. The curved region of the chain plays a key role in the
dynamics, lifting the chain. The chain is in free fall, yet it exhibits whip
dynamics, with one end snapping upward periodically as the chain falls. We
have conducted experiments demonstrating this phenomena.
\end{abstract}


\section{Introduction}

Chains and ropes are among our oldest tools, yet, remarkably, they are still
capable of exhibiting surprising behavior. One example is the Mould effect,
discovered by Stephen Mould \cite{Mould}, where a simple ball chain, pulled
by gravity from a cup, forms an astonishing chain fountain, where the chain
first rises, then turns and falls, forming essentially an inverted catenary 
\cite{Warner1}. Ropes can show similar behavior \cite{rope}. Considerable
effort has gone into unraveling the detailed dynamics in this process \cite%
{warner2, virga, Virga_paradox, Hiroshi, Anghel}. One key ingredient in the
unexpected behavior of chains and ropes is the acquisition of linear
momentum by rigid objects from their environment via hindered rotation \cite%
{acq, Hanna}.

In this paper, we report observations and analysis of a different kind of
unusual behavior: whip dynamics in a freely falling chain. In Section \ref%
{Sect2}, we introduce three facets of simple chain whip dynamics which lays
the foundation of the essence of whip dynamics. In Section \ref{Sect3}, we
study in detail the the dynamics of the free falling chain, both when the
chain is on the pulley, and after liftoff and detachment from the pulley. We
establish the analogy of the free falling chain with simple whip dynamics.
We conclude the paper in Sect.~\ref{Sec_con}.

\section{Prelude: three facets of simple chain whip dynamics}

\label{Sect2}

A traditional whip is a handle to which a flexible strip of leather, of a
length of cord had been attached. Chains instead of leather or cord have
been used as whips, frequently as weapons. Unlike leather or cord whips,
which are typically tapered, chain whips are distinguished by uniform mass
density and low dissipation. Below we discuss three aspects of whip dynamics
which are clearly illustrated by simple 1D models. For simplicity, we ignore
gravity here, and consider energy and momentum conservation.

\subsection{Moving chain with one end arrested}

\begin{figure}[htb]
\centering
\includegraphics[width=.6\textwidth]{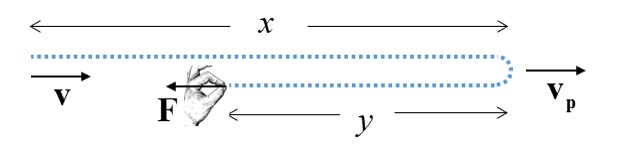}
\caption{Schematic of moving chain with one end arrested}
\label{fig_1}
\end{figure}

We consider an inextensible uniform chain, such as a bicycle chain, of
length $L$, in a straight configuration, moving at constant velocity $%
\mathbf{v}_{0}$ in a direction tangent to the chain. Suppose that a hand
appears at the front of the chain, folds the first link clockwise through
the angle $\pi $, and holds it at rest. (This is essentially how the
cracking of a whip is initiated.) Now the chain is folded, as shown in Fig.~%
\ref{fig_1}, with a moving straight portion on top, and a stationary
straight portion below; the two parts are joined with a (negligibly small)
semicircular section, which we call the 'virtual pulley' (VP), which is also
moving in the same direction.

As time goes on, the top part of the chain, of length $x(t),$ moving with
velocity $\mathbf{v}_{x}(t)$, becomes shorter, while the standing part of
the chain, of length $y(t)=L-x(t)$ at rest, becomes longer; the semicircular
section, which we term the `VP', is moving forward with velocity $\mathbf{v}%
_{p}(t)$. It follows that%
\begin{equation}
\dot{x}=v_{p}-v_{x},
\end{equation}%
and%
\begin{equation}
\dot{y}=v_{p}.
\end{equation}%
and since the chain is inextensible, $\dot{x}+\dot{y}=0$, we have 
\begin{equation}
-\dot{x}=v_{p}=\frac{1}{2}v_{x}.
\end{equation}%
To obtain an explicit expression for the dynamics, conservation of momentum
needs to be considered.

There is tension in the chain which is responsible for changing the momentum
of the chain passing through the `VP'. The speed of the chain relative to
the pulley is $-\dot{x}$. The change of momentum for mass $\rho \Delta L$ in
the VP is $-2(\rho \Delta L)\dot{x}$ which takes place in time interval $%
\Delta t=\Delta L/\dot{x}$ ; this is due to the tension $T$ in the chain at
the two sides of the pulley \cite{warner2}. That is, 
\begin{equation}
T=\rho \dot{x}^{2}.
\end{equation}
The rate of change of momentum in the top part of the chain is\footnote{%
The tension at one point in the chain which changes the momentum of the
attached part does not involve a change in the length of the attached part.} 
$\rho x\dot{v}_{x}=-2\rho x\ddot{x}$, and the rate of change of momentum is
the tension $T$, therefore%
\begin{equation}
\dot{x}^{2}+2x\ddot{x}=0.
\end{equation}%
With initial conditions $x(0)=L$ and $\dot{x}(0)=v_{0}$, the solution is
explicitly given by 
\begin{equation}
x=L(1-\frac{t}{t_{0}})^{\frac{2}{3}},
\end{equation}%
where%
\begin{equation}
t_{0}=\frac{4}{3}\frac{L}{v_{0}}
\end{equation}%
is the time when the chain becomes straight again. We see that momentum
conservation is sufficient to arrive at the equation of motion. This result
can also be obtained by conservation of kinetic energy.

\begin{figure}[tbh]
\centering
(a)\includegraphics[width=.4\linewidth]{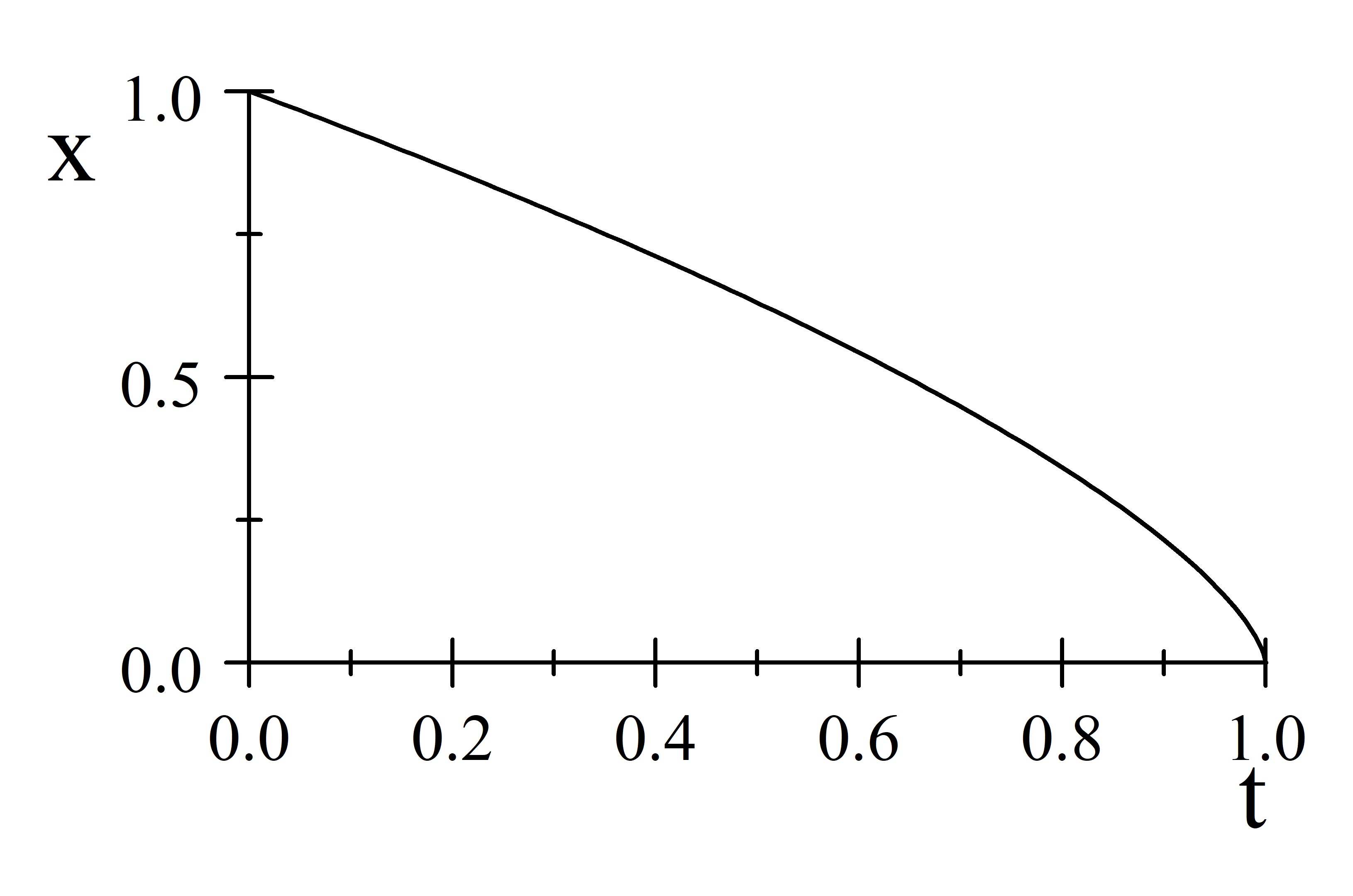} (b)\includegraphics[width=.4%
\linewidth]{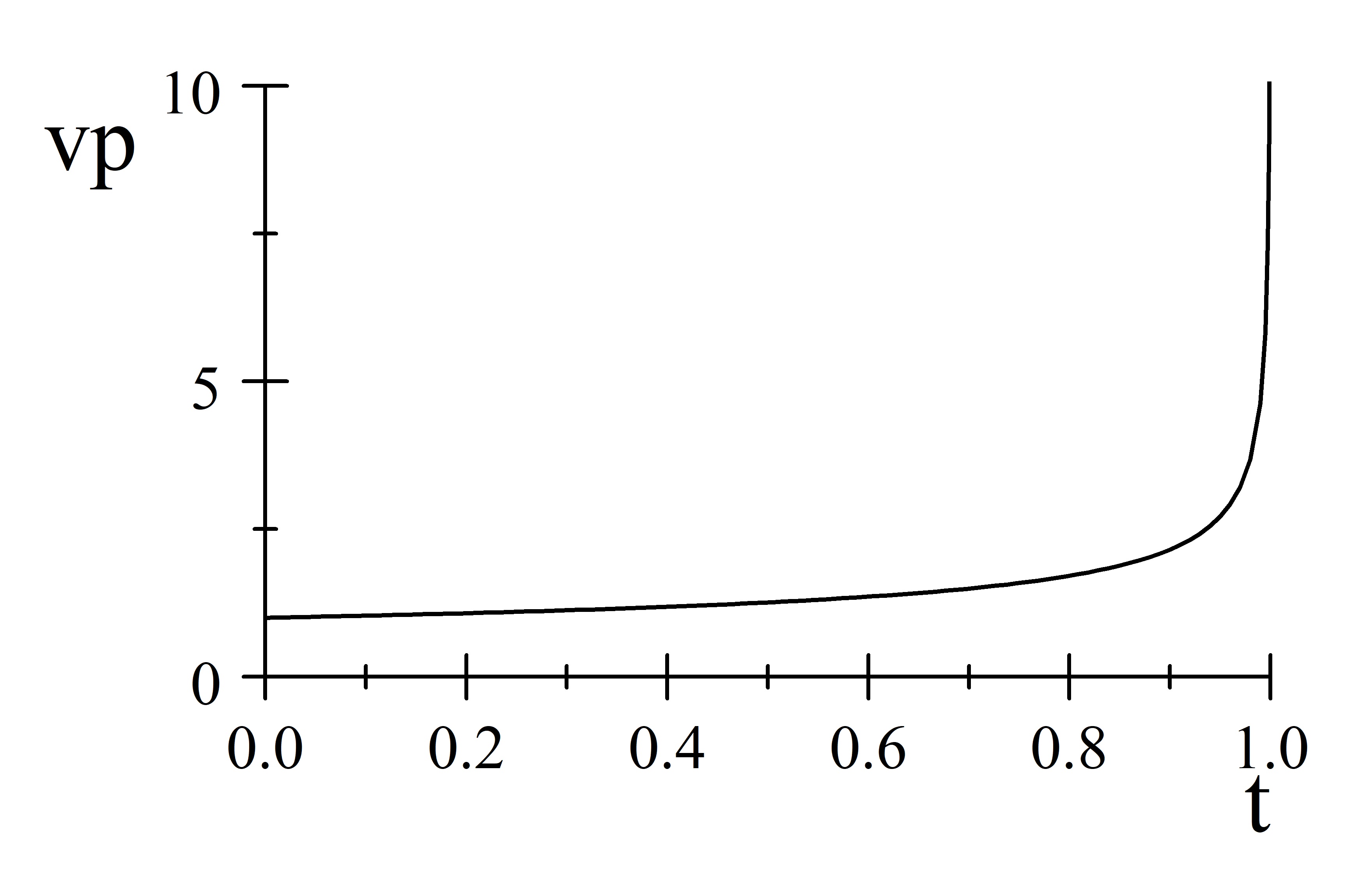}
\caption{(a) The length of the top portion of the moving chain $x$ vs.~$t$.
(b) Speed of virtual pulley $v_{p}$ vs.~$t.$ }
\label{fig_xvst}
\end{figure}

The speed $v_{p}$ of the bend, or the `VP', is then 
\begin{equation}
v_{p}=-\dot{x}=\frac{1}{2}v_{0}(1-\frac{t}{t_{0}})^{-\frac{1}{3}},
\label{eq_vp_1}
\end{equation}%
and both $v_{p}$ and the speed $v_{x}$ of the moving part of the chain
diverge as $t\rightarrow t_{0}$ or as $x\rightarrow 0$. \emph{This is the
essence of the whip dynamics; the conserved kinetic energy becomes localized
in the ever shortening moving part of the whip}. The graphs $x$ and $%
v_{p}/(2L/(3t_{0}))$ as function of $t$ are shown in Fig.~\ref{fig_xvst}. It
is interesting note that $v_{p}$ remains nearly constant until $t\simeq t_{0}
$, and after this, increases rapidly. We see that the velocity of the tip of
an ideal whip diverges.

We now calculate the magnitude of the applied force needed to hold the
bottom chain fixed. Since the net force on the standing part of the chain is
zero, so the force and the tension in the chain at the two sides of the `VP'
are balanced, which gives 
\begin{equation}
F=-T=-\rho \dot{x}^{2}.
\end{equation}%
Substituting Eq.~\eqref{eq_vp_1} for $\dot{x}$, the applied force is 
\begin{equation}
F=-\frac{1}{4}\rho v_{0}^{2}(1-\frac{t}{t_{0}})^{-\frac{2}{3}}.
\end{equation}

The consequence of energy and momentum conservation\footnote{%
We note that the assumption that the chain is lossless has been implicitly
included in the assumption that the tension in the chain on both sides of
the `virtual pulley' is the same.} is that the velocity of the tip an ideal
lossless whip is infinite. It appears in this case that the effect of the
bend in the chain is to constrain the kinetic energy, initially distributed
in the entire chain, to the moving part part on one side of the bend. As the
bend reaches the end of the chain, the energy and momentum densities diverge.

\subsection{Stationary chain with one end pulled}

\begin{figure}[htb]
\centering
\includegraphics[width=.6\textwidth]{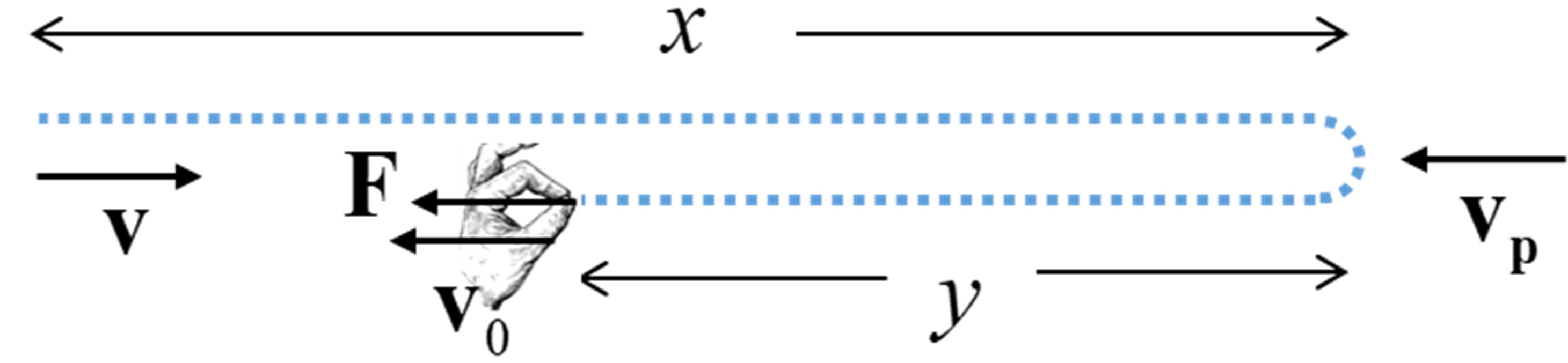}
\caption{Stationary chain with one end pulled}
\label{fig_moving_chain}
\end{figure}

Next, we consider the above scenario viewed from a moving inertial frame of
reference, moving with constant speed $\mathbf{v}_{0}$. In this moving
frame, the chain is initially at rest; then a moving hand appears, and pulls
the stationary chain with velocity $-\mathbf{v}_{0}$. The situation is
illustrated in Fig.~\ref{fig_moving_chain}.

From this new perspective, the only change is that all the velocities have a
constant $-\mathbf{v}_{0}$ added. In this frame, 
\begin{equation}
v_{p}=v_{0}(\frac{1}{2}(1-\frac{t}{t_{0}})^{-\frac{1}{3}}-1)  \label{eq_vp_2}
\end{equation}%
and the `VP' starts to move with velocity $-\frac{1}{2}v_{0}$ at $t=0$, and
speeds up reaching $v_{0}$ at $t=t_{0}$. More interestingly, the top part of
the chain has velocity $v_{x}$ given by 
\begin{equation}
v_{x}=v_{0}((1-\frac{t}{t_{0}})^{-\frac{1}{3}}-1).
\end{equation}

As soon as the bottom part of the chain is pulled to the left with constant
velocity $v_{0}$, remarkably, the entire top part of the chain moves and
accelerates to the right! Although not easy to observe on a tabletop due to
friction, the effect can clearly be seen. Applying a force in one direction
to one part of the chain, gives rise to an equal and opposite force acting
on the other part. Forces are unchanged when viewed from inertial frames,
and the force accelerating the top part is just the applied force. The
folded chain appears to guide tensile stress; the stress, initiated by the
pull, traverses the bend and the sample, and `pulls' it, causing the far end
to move in the opposite direction from the external pull.

\subsection{Chain dynamics with no external force}

In the third scenario, we consider a chain in space, at $t=0$, folded
exactly in half. The folded chain is horizontal, the fold is at the right.
The top half is moving to the right, with velocity $v_{x}(t)$, the bottom
half is moving to the left, with velocity $v_{y}(t)$, as shown in Fig.~\ref%
{fig_fold_chain}. The center of mass of the chain is at rest, since there
are no external forces acting on the chain.

\begin{figure}[htb]
\centering
\includegraphics[width=.6\linewidth]{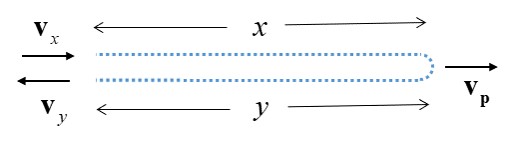}
\caption{Moving chain without external force}
\label{fig_fold_chain}
\end{figure}

As before, we have that 
\begin{equation}
\dot{x}=-v_{x}+v_{p},  \label{eq_3_1}
\end{equation}%
and%
\begin{equation}
\dot{y}=v_{y}+v_{p}.  \label{eq_3_2}
\end{equation}%
Since $\dot{x}+\dot{y}=0$, we have%
\begin{equation}
v_{p}=\frac{1}{2}(v_{x}-v_{y}).
\end{equation}

Since there are no external forces acting and the initial total momentum is
zero, we have%
\begin{equation}
\rho xv_{x}-\rho yv_{y}=0.  \label{eq_3_con}
\end{equation}%
Substituting Eqs.~\eqref{eq_3_1} and \eqref{eq_3_2}, and $x+y=L$ into Eq.~%
\eqref{eq_3_con}, gives 
\begin{equation}
v_{p}=\dot{x}(\frac{2}{L}x-1).  \label{eq_vp_3}
\end{equation}

The tension in the chain at the two sides of the `VP' is, as before%
\begin{equation}
T=\rho \dot{x}^{2}
\end{equation}%
and force balance on the top part gives%
\begin{equation}
\rho x\dot{v}_{x}=\rho \dot{x}^{2}.  \label{eq_3_4}
\end{equation}%
From \eqref{eq_3_1}, we have $\dot{v}_{x}=\dot{v}_{p}-\ddot{x},$ and from %
\eqref{eq_vp_3} 
\begin{equation}
\dot{v}_{p}=\ddot{x}(\frac{2}{L}x-1)+\frac{2}{L}\dot{x}^{2}.
\end{equation}%
Substitute those relations in Eq.~\eqref{eq_3_4}, we arrive at 
\begin{equation}
\ddot{x}x(\frac{x}{L}-1)+\dot{x}^{2}(\frac{x}{L}-\frac{1}{2})=0.
\label{keyeq}
\end{equation}%
This is the equation of motion. After some manipulation, this can be solved
to give an implicit relation between $x$ and $t,$ 
\begin{equation}
t=\frac{L}{2v_{0}}((1-2\frac{x}{L})\sqrt{\frac{x}{L}(1-\frac{x}{L})}+\frac{1%
}{2}\sin ^{-1}(1-2\frac{x}{L})).  \label{expl}
\end{equation}%
Furthermore, we obtain 
\begin{equation}
\dot{x}=\frac{-v_{0}}{2\sqrt{\frac{x}{L}(1-\frac{x}{L})}}.
\end{equation}%
It follows that $\dot{x}$ diverges as $x\rightarrow 0$, so do $v_{p}$ and $%
v_{x}$. A plot of $x$ and $v_{p}$ as function of time are shown in Fig.~\ref%
{fig_xvst_2}.

\begin{figure}[tbh]
\centering
(a)\includegraphics[width=.4\linewidth]{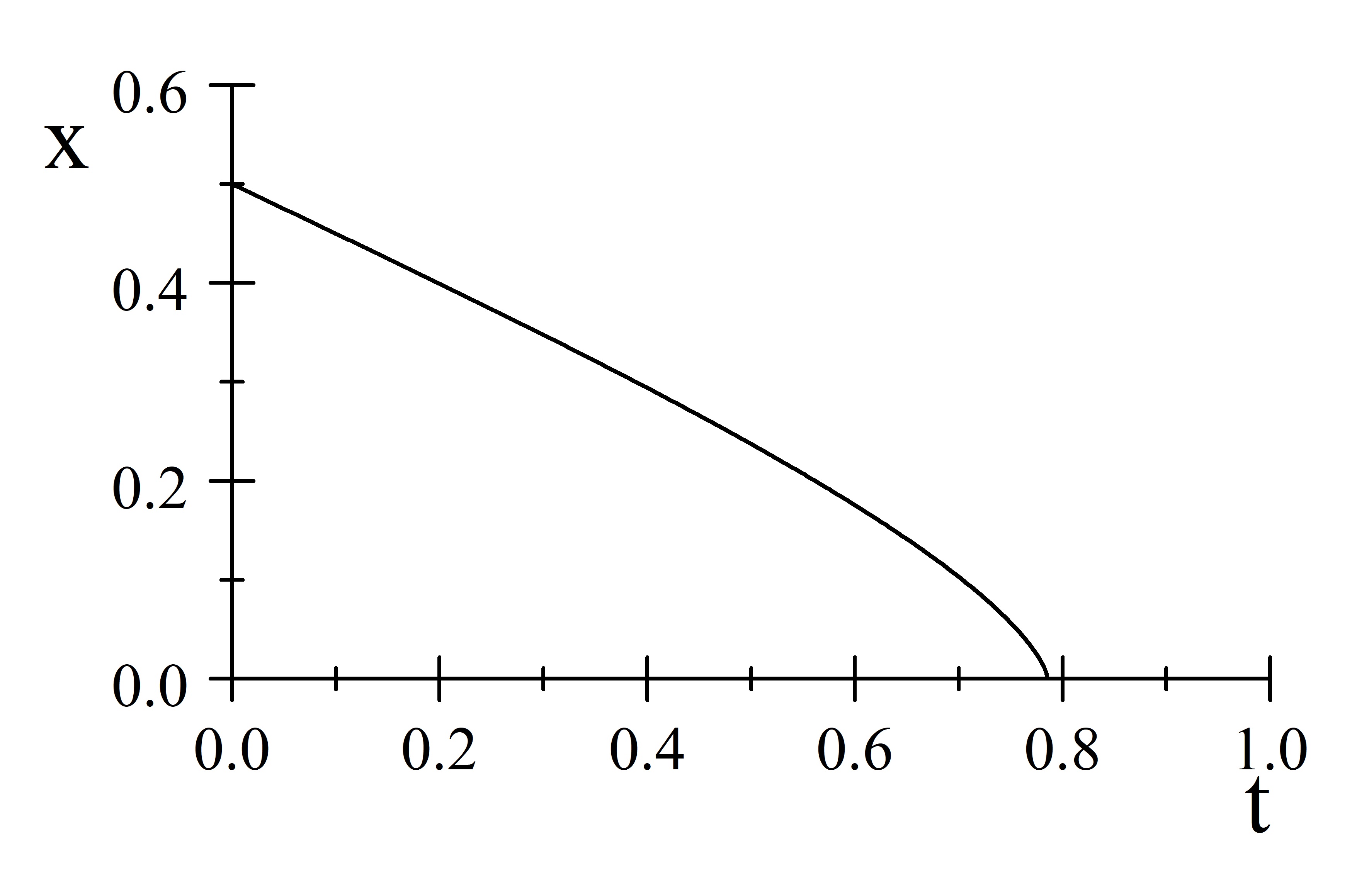} (b) %
\includegraphics[width=.4\linewidth]{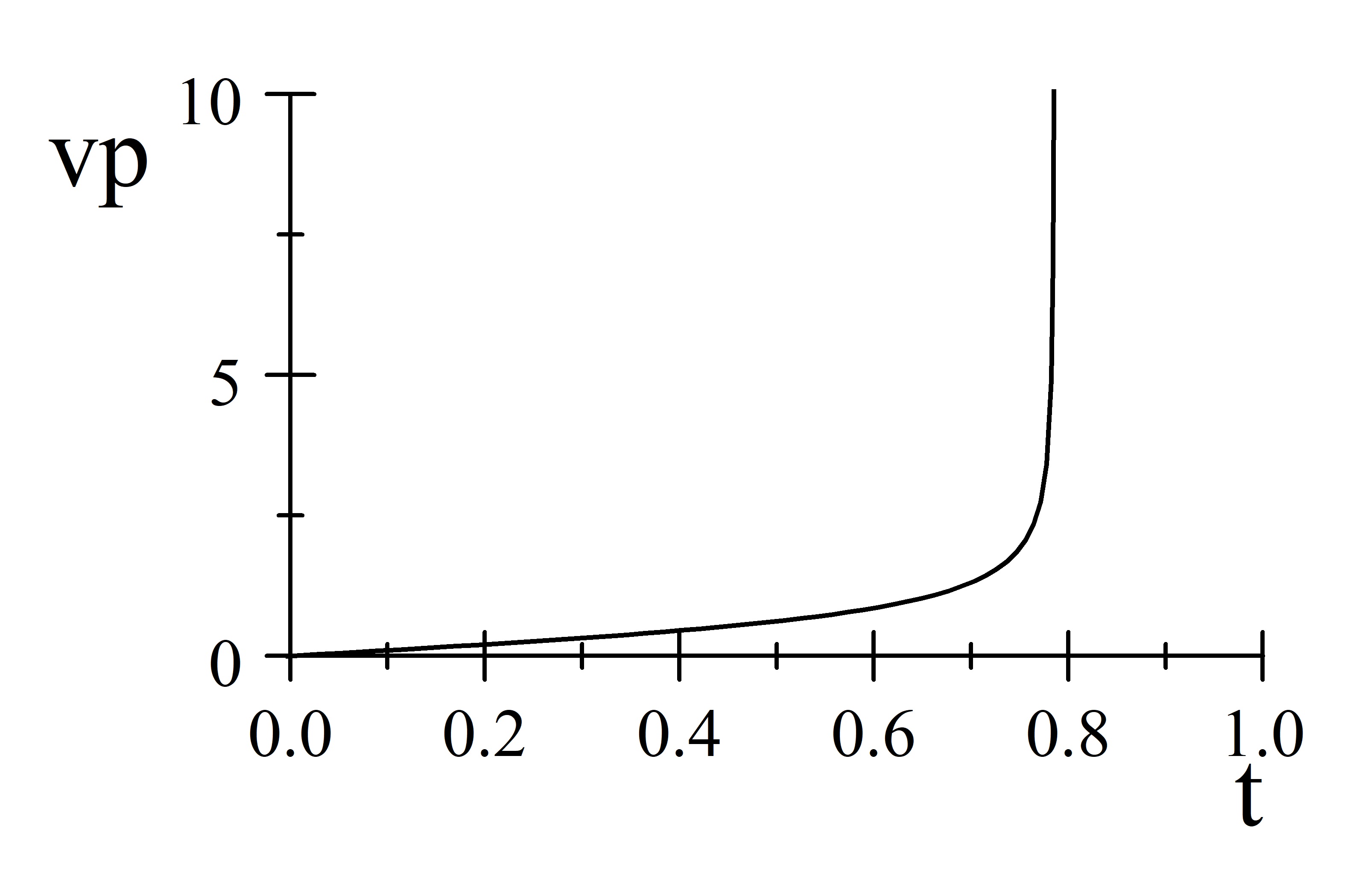}
\caption{(a) The length of the top portion of the moving chain $x$ vs.~$t$.
(b) velocity of the virtual pulley vs.~$t$.}
\label{fig_xvst_2}
\end{figure}

\begin{figure}[tbh]
\centering
\includegraphics[width=.4\linewidth]{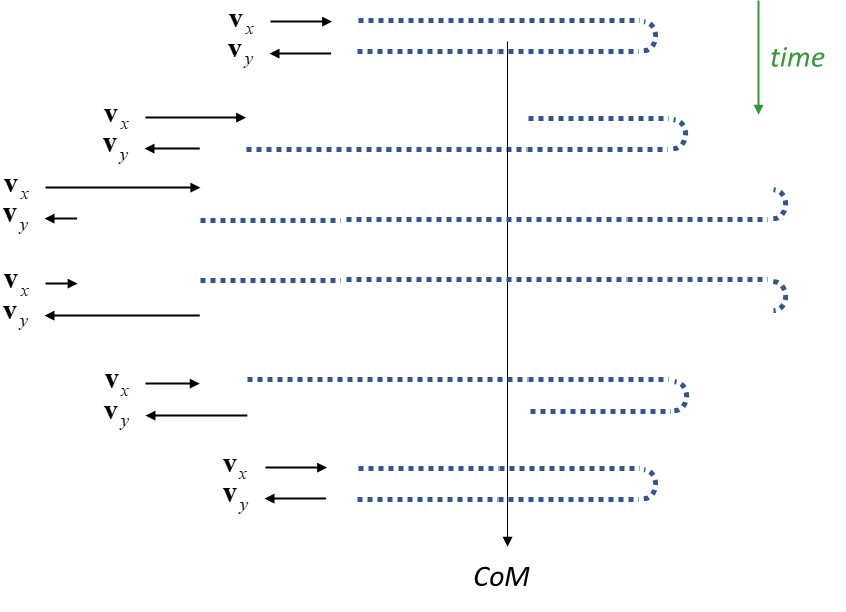}
\caption{A schematic showing the periodic dynamics of a moving chain
dynamics. The center of mass is translating in the direction of arrow, and
the system has a nonvanishing constant angular momentum.}
\label{fig_per}
\end{figure}

Since the initial momentum is zero, when the chain is straight, chain comes
to rest, except for the last piece; all the kinetic energy of the entire
chain ends up in the tip! The tip will rotate and move rapidly towards the
left as shown in Fig.~\ref{fig_per}. The part of the chain now at the bottom
will move towards the left and slow down, eventually reaching the
conformation identical to the initial one. In this simple model, the chain
will continue this periodic motion.

In this prelude, we have shown that a folded chain with moving segments can
concentrate all of its kinetic energy at one point - at one of the tips -
with or without an external force. The folded chain also appears to guide
tensile stress around the bend, and give rise to motion there opposite to
the direction of the force causing the motion. If allowed to continue, the
concentrated energy and momentum densities will spread out, and the process
will continue, without dissipation, indefinitely.

We believe these aspects constitute the dynamics normally associated with
whips. Essentially the same phenomena are exhibited by other flexible strips
of material; however, energy losses associated with dissipative processes
which are often present can mask the behavior.

\section{Freely falling chain}

\label{Sect3}

\subsection{Chain on a fixed pulley}

\begin{figure}[htb]
\centering
\includegraphics[width=.2\textwidth]{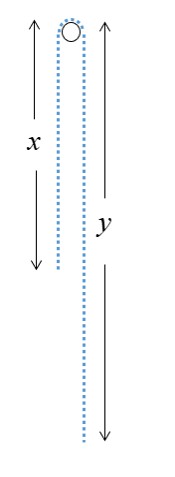}
\caption{Chain on a fixed pulley}
\label{fig_falling_chain}
\end{figure}

We next consider the behavior of a chain of length $L$ and mass density $%
\rho $ on a fixed pulley in the presence of gravity. This is illustrated in
Fig.~\ref{fig_falling_chain}. The length of chain on the left is $x$, and at
time $t=0$, it is moving up with velocity $v_{0}$, while the length of the
chain on the right is $y=L-x$, and it is moving down. We assume the radius
of the pulley is arbitrarily small. The chain is pulled down by gravity, and
the pulley is supported by an upward force $F(t)$. The tension in the chain
at the pulley is $T(t)$. The chain is moving with speed $v(t)=-\dot{x}$. The
net force on the pulley is 
\begin{equation}
F-2T=-2\rho \dot{x}^{2}=-2\rho \dot{y}^{2},  \label{s1}
\end{equation}%
where the right hand side is the rate of change of momentum of the chain on
the pulley. 

Consider the chain on the left. The equation of motion is 
\begin{equation}
T-\rho xg=-\rho x\ddot{x}.  \label{s2}
\end{equation}
Combining Eqs.~\eqref{s1} and \eqref{s2} to get 
\begin{equation}
\rho x\ddot{x}+\rho \dot{x}^{2}-\rho xg=-\frac{1}{2}F.
\end{equation}%
On the right side of the chain, we have%
\begin{equation}
\rho yg-T=\rho y\ddot{y},
\end{equation}%
and proceeding similarly, we obtain 
\begin{equation}
-\rho (L-x)\ddot{x}+\rho \dot{x}^{2}-\rho (L-x)g=-\frac{1}{2}F.
\end{equation}

It is convenient, for simplicity, to use dimensionless variables. We measure
all lengths in units of $L$, time in units of $\sqrt{L/2g}$ and force in
units of $4\rho Lg$. We then have%
\begin{equation}
x\ddot{x}+\dot{x}^{2}-\frac{1}{2}x=-f,  \label{fif}
\end{equation}%
and 
\begin{equation}
-(1-x)\ddot{x}+\dot{x}^{2}-\frac{1}{2}(1-x)=-f.
\end{equation}%
Subtracting these, we get 
\begin{equation}
\ddot{x}-x+\frac{1}{2}=0.  \label{bef}
\end{equation}%
Eq.~\eqref{bef} is the equation of motion when the chain is on the pulley.
This can be solved at once; the solution is 
\begin{equation}
x=\frac{1}{2}+Ae^{t}+Be^{-t},
\end{equation}%
where $A$ and $B$ depend on initial conditions $x_0$ amd $v_0$ as 
\begin{equation}
A=\frac{1}{2}((x_{0}-\frac{1}{2})-v_{0}),
\end{equation}%
and%
\begin{equation}
B=\frac{1}{2}((x_{0}-\frac{1}{2})+v_{0}).
\end{equation}%
As expected, if $A<0$, $\dot{x}<0$, and the speed of the chain speed
increases nearly exponentially.

From Eq.~\eqref{fif} the force $f$ on the pulley can be readily calculated, 
\begin{equation}
f=\frac{1}{4}\left( 1-8A^{2}e^{2t}-8B^{2}e^{-2t}\right) .
\end{equation}%
As the chain speeds up, the force $f$ exerted on the pulley by the chain
decreases with time and reaches zero at time $t^{\ast }$ given by 
\begin{equation}
t^{\ast }=\frac{1}{2}\ln \left( \frac{1}{16A^{2}}\left( \sqrt{1-256A^{2}B^{2}%
}+1\right) \right).
\end{equation}%
At that time, the chain rises and lifts off the pulley! We again regard the
bend in the chain again as a `VP'.

A relevant question is the length of the upward moving part of the chain at
liftoff. A straightforward calculation gives 
\begin{equation}
x=\frac{1}{2}-\frac{1}{2}\sqrt{\frac{1}{2}+2(x_{0}-\frac{1}{2}%
)^{2}-2v_{0}^{2}}.
\end{equation}%
We note that the quantity under the square root needs to be nonnegative, and
that we must have $x<x_{0}$. These lead to 
\begin{equation}
(\frac{1}{2}-x_{0})^{2}+v_{0}^{2}<\frac{1}{4}.
\end{equation}%
This must be satisfied for liftoff. In particular, starting from rest, with $%
x_{0}<\frac{1}{2}$ should always work (with a frictionless pulley!);
starting at $x_{0}=1/2$ should always work if $v_{0}<1/2$.

\subsection{After liftoff: the freely falling chain}

After liftoff, the force from pulley vanishes, that is, $f=0$. Proceeding as
before, we now arrive at, for the chain on the left, in dimensionless units, 
\begin{equation}
x\ddot{x}+\dot{x}^{2}-x\dot{v}_{p}=0,
\end{equation}%
and for the chain on the right, 
\begin{equation}
-(1-x)\ddot{x}+\dot{x}^{2}-(1-x)\dot{v}_{p}=0.
\end{equation}%
%
%
%
We can eliminate $\dot{v}_{p}$, and in the presence of gravity, we have the
equations of motion%
\begin{equation}
2\ddot{x}x(1-x)+\dot{x}^{2}(1-2x)=0,
\end{equation}%
\begin{equation}
\dot{v}_{p}=\dot{x}^{2}\frac{1}{2x(1-x)}-\frac{1}{2}.
\end{equation}%
Recall that without gravity, we have%
\begin{equation}
2\ddot{x}x(1-x)+\dot{x}^{2}(1-2x)=0,
\end{equation}%
\begin{equation}
\dot{v}_{p}=\dot{x}^{2}\frac{1}{2x(1-x)}.
\end{equation}%
The only difference is the term $-1/2$ in the equation for $\dot{v}_{p}$,
this is just the dimensionless acceleration of gravity. So we should see
essentially the same phenomenon without gravity (say for the chain on a
frictionless table) except the falling of the center of mass. Without
gravity, we don't have the initial conditions required for takeoff.

We also have the result that the height $h$ of the center of mass and the
height $h_{p}$ of the center of pulley are related;%
\begin{equation}
h=-x^{2}+x+h_{p}-\frac{1}{2}.
\end{equation}%
Setting $h=0$, we have $h_{p}(t)$ once we have $x(t)$ 
\begin{equation}
h_{p}=x^{2}-x+\frac{1}{2}.
\end{equation}%
We also have 
\begin{equation}
v_{p}=(2x-1)\dot{x}.
\end{equation}%
Therefore if $x<0.5$ and $\dot{x}<0$, the `VP' rises.

\subsection{Experimental Results}

We carried out simple experiments to observe the dynamics of chains,
initially on a pulley. Our goal was to compare model predictions from
Sects.~3.1 and 3.2 above with experimental observations. Due to the
challenging requirements of making reliable measurements on the length and
timescales involved, we only carried out simple observations at this
preliminary stage of work.

Since the timescale of our experiments is $\sqrt{L/g}$, a length of $10m$
corresponds to $1s$. To observe the development of the dynamics and the
transition between the different regimes, we chose to work in an abandoned
stairwell of Kent State University Library. The building has 12 floors, as
well as a basement. The distance from the 10th floor, which was our station,
to the basement floor was $42.7m$. For our experiments, we used $3/16"$
nickel-plated steel ball-chains from McMaster-Carr, with mass density $\rho
=30g/m$ and maximum curvature $1.7cm^{-1}$. Instead of a pulley, we used a
polished steel pipe with $2cm$ o.d.

\begin{figure}[tbh]
\centering
\includegraphics[width=.6\linewidth]{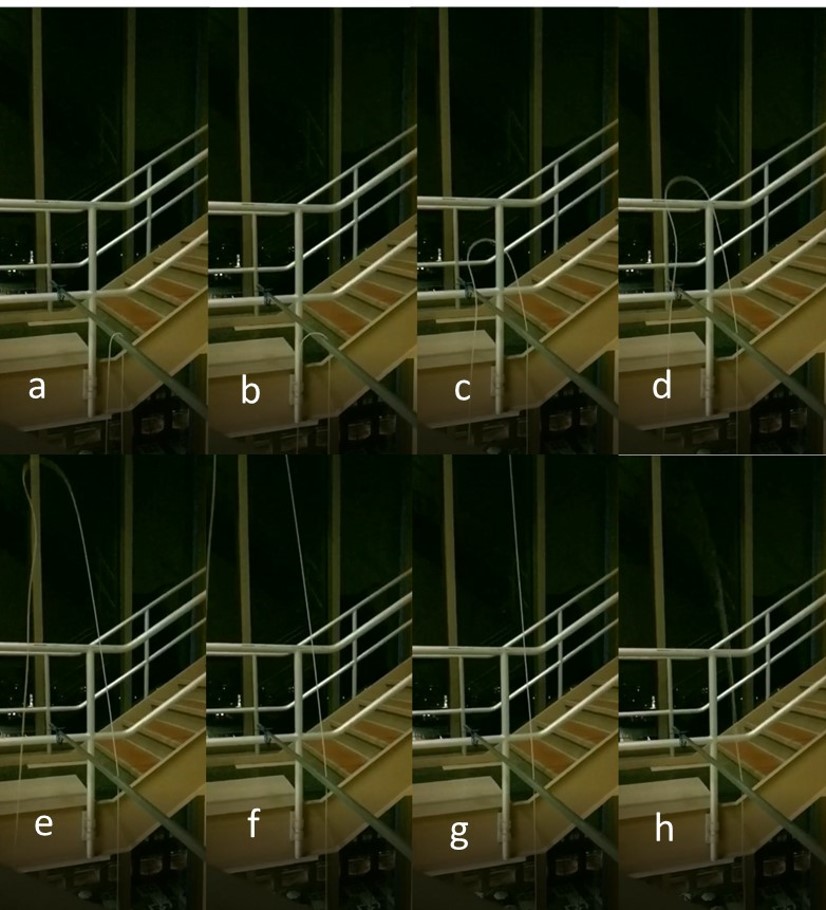}
\caption{Sequence of stills from the video showing the chain dynamics. The
times are: a=0s,b=2s,c=3s,d=4s,e=5s,f=6s,g=7s and h=8s from start of the
video.}
\label{fig_exp}
\end{figure}

In the experiment described here, we used a $L=61m$ chain. The chain was
draped over the supporting steel pipe, with equal lengths on each side of
the bar. With the chain at rest, we pulled the chain on one side of the bar
downward, by hand, to make one side longer. At one point in this process,
chain started slip on the supporting pipe, with the longer side moving
downward, and the shorter side moving upward. We took a video of the chain
over the bar; we started the video once the chain was moving with
appreciable speed. The speed of the chain increased more and more rapidly in
time, making considerable noise as it passed over the pipe. As the chain
speeded up, but still in contact with the pipe, the curvature of the `VP', that is, of the top of the chain, decreased. About $2s$ after the
video was started, the chain lifted off the pipe, and the VP rose rapidly
and accelerated upward, while its curvature decreased significantly. We
estimate that the in the first second after liftoff, VP rose $1m$, in the
second second it rose $2m$, and continued to rise (off camera) for another $%
2 $ seconds. At that time, we heard a sound, which was consistent with the
chain hitting the ceiling at the top of the stairwell. We found that the
chain had indeed hit the ceiling, leaving marks and embedded pieces of chain
in the fiber ceiling tiles. Subsequent experiments produced similar results.
We estimate that the VP in the video rose in excess of $7m$ to hit the
ceiling. Other experiments using 4" PVC pipe for support, and $1/4"$
ball-chains we observed even greater heights. We attach stills from our
video in Fig.~\ref{fig_exp}, which shows the chain supported by the pipe,
lifting off the pipe, the VP accelerating as it rises to the ceiling of the
stairwell. The last frame shows the chain falling back.

\section{Summary}

\label{Sec_con}

We have considered essential aspects of a lossless whip; these are the
concentration of the energy and momentum densities into a single point at
the tip, and the subsequent spreading out of these again throughout the
chain. In the absence of losses, this process repeats indefinitely. We have
also argued that if a straight chain at rest is folded in half, and if then
one end is pulled away from the fold, the VP, other end will move towards it.

We have considered a chain with free ends hanging from a pulley which is
supported from below. If the system is unbalanced, the chain speed grows
nearly exponentially, and for a given range of parameters, at a given time,
the chain lifts off the pulley. After this time, the dynamics of the chain
is the same as that of a whip; although the center of mass of the chain is
in free fall. Due to the whip dynamics, one end of the chain snaps upward,
and if the chain is lossless, this snapping of alternating ends is expected
to continue indefinitely.

We have carried out experiments where long slender ball-chain is supported
by a horizontal metal rod, so that one half of the chain hangs on one side,
and the other half hangs on the other side. If the chain is then positioned
so that there is more chain on one side than the other and the chain is
brought into motion, the chain will begin to slip with increasing speed,
with the longer part of the chain moving downward and the shorter part
moving upward. At a critical speed, the chain lifts off the supporting bar,
with the curved region where the chain changes direction, that is, with the
`VP', accelerating rapidly upward. In our experiments, the top part of the
chain rose 20' above the bar and came into collision with the ceiling tiles.
Although the center of mass of the chain is freely falling after lift-off,
the chain exhibits whip dynamics with the upward moving part snapping; in
our case, the tip of the chain disintegrated into numerous small fragments.
The initiation of lift-off of a chain from a pulley has been observed and
examined by \cite{audoly}; however, they imposed a constant acceleration on
the chain, and only a very small fraction of the chain (less than the pulley
circumference), at the end of the chain, lifted off the pulley. Here, we see
fully developed liftoff, far from the end of the chain, with the upward and
downward moving parts of the chain nearly parallel. In addition to
exhibiting fascinating dynamics, this mechanism may be useful for imparting
large upward velocities to objects at the end of the chain, and thus for
launching projectiles.

In this paper, using 1D dynamics, we have assumed that the curvature is a
constant at the fold, which is not the case, as experiments show. This will
be a topic for a future study.

A video showing the dynamics of the falling chain, together with additional
information, is available at \href{http://mpalffy.kent.edu/chain}{%
http://mpalffy.kent.edu/chain}.

\section*{Acknowledgments}

We are grateful for illuminating discussions with M. Warner (University of
Cambridge) and H. Yokoyama (Kent State University), and to R. Bammerlin and
Kent State Libraries for access to the building and stairwell. This work was
supported by the US Office of Naval Research [ONR 00014-18-1-2624]. Results
from this paper has been presented at the Minisymposium MS 125 ``Liquid
Crystals, Elastomers and Beyond: In Memory of Mark Warner" at 2022 Annual
Meeting of Society for Industrial and Applied Mathematics.

\end{document}